\documentstyle[12pt,epsf]{article}

\def\singlespace 
{\smallskipamount=3.75pt plus1pt minus1pt
\medskipamount=7.5pt plus2pt minus2pt
\bigskipamount=15pa plus4pa minus4pt \normalbaselineskip=12pt plus0pt
minus0pt \normallineskip=1pt \normallineskiplimit=0pt \jot=3.75pt
{\def\smallskip {\vskip\smallskipamount}} {\def\medskip
{\vskip\medskipamount}} {\def\bigskip {\vskip\bigskipamount}}
{\setbox\strutbox=\hbox{\vrule height10.5pt depth4.5pt width 0pt}}
\parskip 7.5pt \normalbaselines} 

\def\middlespace
{\smallskipamount=5.625pt plus1.5pt minus1.5pt \medskipamount=11.25pt
plus3pt minus3pt \bigskipamount=22.5pt plus6pt minus6pt
\normalbaselineskip=22.5pt plus0pt minus0pt \normallineskip=1pt
\normallineskiplimit=0pt \jot=5.625pt {\def\smallskip
{\vskip\smallskipamount}} {\def\medskip {\vskip\medskipamount}}
{\def\bigskip {\vskip\bigskipamount}} {\setbox\strutbox=\hbox{\vrule
height15.75pt depth6.75pt width 0pt}} \parskip 11.25pt
\normalbaselines} 

\def\doublespace 
{\smallskipamount=7.5pt plus2pt minus2pt \medskipamount=15pt plus4pt
minus4pt \bigskipamount=30pt plus8pt minus8pt \normalbaselineskip=30pt
plus0pt minus0pt
\normallineskip=2pt \normallineskiplimit=0pt \jot=7.5pt
{\def\smallskip {\vskip\smallskipamount}} {\def\medskip
{\vskip\medskipamount}} {\def\bigskip {\vskip\bigskipamount}}
{\setbox\strutbox=\hbox{\vrule height21.0pt depth9.0pt width 0pt}}
\parskip 15.0pt \normalbaselines}

\begin{document}
\bibliographystyle{unsrt}

\begin{flushright} 

June, 1996

hep-ph/9606447

\end{flushright}

\vspace{3mm}

\begin{center}

{\Large \bf Magnitude of R-parity violation in String Inspired GUTs}\\ 

\vspace{10mm}
{\bf M. Bastero-Gil$^a$, B. Brahmachari$^b$\footnote{ Address after 1st 
October 1996: Department of Physics, University of Maryland, College
Park, MD 20742, USA.} and R.N. 
Mohapatra$^c$\footnote{Work supported by the
 National Science Foundation Grant \#PHY-9119745 and a Distinguished
Faculty Research Award by the University of Maryland for the year 1995-96.}}\\
\vskip 8mm
{\it $^a$ Scoula Internazionale Superiore di Studi Avanzati,\\ 34013 Trieste, Italy.}\\
\vskip 3mm
{\it $^b$ International Centre for Theoretical Physics, \\34100 Trieste, Italy.}\\
\vskip 3mm
{\it{$^c$ Department of Physics, University of Maryland.}}\\

{\it{ College Park, MD 20742 }}

\end{center}

\vspace{3mm}

\begin{center}
{\bf Abstract}
\end{center}
\vspace{1mm}

The nature of R-parity violating interactions in two classes of string 
inspired supersymmetric grand unified theories (SISUSY GUT), based on the 
gauge groups $SO(10)$ 
(and its subgroup $SU(2)_L\times SU(2)_R\times SU(4)_c\equiv G_{224}$)
as well as $[SU(3)]^3$, are discussed and their strengths are related to the 
ratio of symmetry breaking scales present in the 
model. We first argue that for the R-parity violating couplings
$\lambda_{R\!\!\!/}$ to be suppressed to the desired level, the $B-L$ local 
symmetry must break at an intermediate scale $M_{B-L}$ since
$\lambda_{R\!\!\!/} =M_{B-L}/M_{GUT}$. We then 
construct scenarios where such intermediate scales arise being consistent with 
gauge coupling unification in a two-loop renormalization group study. 
In the resulting $SO(10)$ models, higher-dimensional-operator-induced
R-parity violating couplings are potentially large (except in one case),
and are therefore
inconsistent with limits on the proton life time unless 
the couplings associated with the higher dimensional terms
are fine tuned to very small values.
However, the $[SU(3)]^3$ 
and $G_{224}$ models can be consistent if a certain class of quark-lepton 
coupling in the superpotential is forbidden by a discrete symmetry 
(unrelated to R-parity). An interesting prediction of these models is
neutron-anti-neutron oscillation with observable strength. 

\newpage
\doublespace
\section{Introduction}

The next step beyond the standard model is now widely believed to
be the supersymmetric version of the standard model (to be called the MSSM) 
with a supersymmetry breaking scale around or less than the TeV scale. While
this approach solves the problem of the Higgs mass and has the potential to
provide a radiative origin for the electroweak symmetry breaking, it
has a troubling conceptual problem, having to do with the fact that
it allows for lepton and baryon number violating interactions with
arbitrary strengths. These are the so-called R-parity violating
interactions. There exist very stringent upper limits
on the various R-parity violating couplings \cite{rv} which range
anywhere from $10^{-4}$ to $10^{-12}$ depending on the type of
selection rules they break. The most restrictive of them are on the
products of $\Delta B=1$ and $\Delta L=1$ couplings which follow
from proton decay experiments\cite{hinchliffe,vissani}; there are also 
limits which follow from neutron-anti-neutron oscillation
\cite{zw,goity} as well as from considerations
of cosmological baryon asymmetry if additional assumptions are made
regarding the primary origin of the asymmetry\cite{campbell}.
Perturbative unitarity arguments \cite{pertur,rpvsoft}
also put upper bounds on the
R-parity violating couplings.
Since the main reason for believing in supersymmetry
is that it improves the naturalness of the standard model, it will be 
awkward to assume that the MSSM carries along with it this
``baggage" of fine-tuned couplings without any fundamental reason. 

The general attitude to this problem is
that when the MSSM is extrapolated to higher scales, new symmetries
will emerge which either forbid the R-parity violating couplings or
suppress it in a natural manner. A concrete proposal in this direction
made some time ago \cite{rm1} and followed up in several 
papers \cite{martin} is that at higher energies the gauge symmetry
becomes bigger and includes $B-L$ as a subgroup\footnote{A different class of
models where intermediate scale symmetry is the Peccei-Quinn symmetry
has recently been discussed by Tamvakis \cite{tamvakis}.}. 
The $B-L$ gauge group
also is important in understanding the smallness of the neutrino mass;
therefore this is not a completely new symmetry custom-designed
only to solve the R-parity problem. It is easy to see that in the 
symmetric phase of a theory containing $B-L$ local symmetry, R-parity
is conserved since $R=(-1)^{3(B-L)+2S}$. This however is not the
end of the story since the $B-L$ must be a broken symmetry at low
energies. If the $B-L$ symmetry is broken by the vev of a scalar field
which carries odd $B-L$, then R-parity is again broken at low 
energies\cite{rm1,martin}. Examples of  theories where R-parity is
broken by such fields could be the string inspired $SO(10)$ and $[SU(3)]^3$
models. On the other hand there are also many theories
where $B-L$ is broken by fields with even $B-L$ values \cite{kuchi,rm2,leemoh}.
In these models, R-parity remains an exact symmetry, as is required if
supersymmetry has to provide a cold dark matter particle. It remains to
be seen whether these latter class of models can arise from some higher
level compactification of superstring theories.

In this paper we focus on the first class of theories since it has been
shown that they can arise from string models in different compactification
schemes\footnote{Whether one can obtain models with three chiral generations
in these models is not settled; however, this is not relevant 
for our discussion of R-parity breaking.}. 
In this class of theories, R-parity breaking interactions arise
once the $B-L$ symmetry is broken; as a result, the strength of R-violating
interactions depends on the scale at which $B-L$ is broken.
To suppress them to the desired level, $B-L$ breaking must
occur at an intermediate scale \cite{lee,brahm} rather 
than at the GUT scale as is true in most current discussions of SUSY GUT.
The two models we will consider are (i) an $[SU(3)]^3$ model
and (ii) an $SO(10)$ model (and its subgroup $SU(2)_L\times SU(2)_R\times
SU(4)_c$ to be denoted $G_{224}$). We will show that
in both cases our scenarios are consistent with gauge coupling 
unification. The $SO(10)$ models have been discussed by us 
recently \cite{lee,brahm,brahm1}; the $[SU(3)]^3$ scenario is a new 
one, which also has the interesting feature that the GUT scale for it 
coincides with the string scale as in the SO(10) case considered 
previously \cite{brahm1}. Two generic scenarios emerge from this 
study: one in which the intermediate
scale, $M_{B-L}$ is of order $10^{6}$ GeV and a second one where the $M_{B-L}$ 
is of order $10^{13}$ GeV. Clearly the first one 
is consistent with all known constraints on R-parity violating couplings,
whereas the second case raises question about the viability of the
$SO(10)$ model. However, in the $[SU(3)]^3$ and $G_{224}$ cases
 if an additional
discrete symmetry is imposed to forbid certain couplings in the 
superpotential, the model leads to $N-\bar{N}$ oscillation with
observable strength.

Let us begin by writing down the general structure of R-parity violating
interactions in the MSSM:
\begin{eqnarray}
W_{RP}=\lambda_{ijk} L_i L_j E^c_k+\lambda'_{ijk} Q_i L_j D^c_k +
\lambda''_{ijk} U^c_i D^c_j D^c_k
\label{eq1}
\end{eqnarray}

As already mentioned, there are stringent upper limits on the 45 coupling
parameters in Eqn. (\ref{eq1}). We would therefore like to look for theories where
the smallness of these parameters will arise naturally. As already mentioned,
we will be interested in grand unified theories which contain $B-L$ as a
subgroup. We will therefore consider only the SO(10),
$G_{224}$ and $[SU(3)]^3$
theories.  For detailed study of these models, we need to know their matter
and Higgs content. Here we will be guided by the predictions of the superstring
models compactified fermionically \cite{lykken,dienes}.
It turns out that complete breakdown of the gauge symmetry in these cases
automatically imply that R-parity, which is an exact symmetry  above the
GUT scale breaks down. Our goal will be to study the prediction of the
strength R-parity violating interactions in these models consistent with
the idea of gauge coupling unification. Let us proceed to study these models
in turn.

\section{Spontaneous breaking of R-parity in string inspired
GUT models and need for an intermediate scale.}


\noindent{\it SO(10) case:}


As is well-known, the matter fields belong to the spinor {\bf 16}-dimensional
representations whereas the Higgs fields will belong to {\bf 45}, {\bf 54},
{\bf 16}+$\overline{\bf 16}$, {\bf 10}-dim 
representations as is suggested by recent studies of level
two models\cite{lykken}. It has recently been shown\cite{dienes} that
these representations are indeed the only ones that appear in free-field
heterotic string models regardless of the affine level at which $SO(10)$
is realized. The symmetry breaking in these models is achieved as
follows: The vev of the {\bf 45} and {\bf 54}-dim fields 
break the $SO(10)$ symmetry down to $SU(3)_c\times SU(2)_L\times SU(2)_R\times
U(1)_{B-L}$ which is broken down to the standard model by the ${\nu_H^c}$
component of ${\bf 16_H}+\overline{\bf 16_H}$ acquiring 
vevs at the intermediate 
scale. The question we now ask is what is the origin of R-parity violating 
terms at low energies. If we restrict ourselves only to renormalizable 
interactions, then such terms can arise from interactions of type 
$f_m {\bf 16}_m {\bf 10}_H {\bf 16}_H$,  where
subscripts $m$ and $H$ stand for matter and Higgs type fields when we 
substitute ${\nu_H^c}$ vevs. In the language of the MSSM fields, the 
induced operator looks like $L_mH_u$ with a mass coefficient of order 
$f_mv_{BL}$. This term will lift both the L and the $H_u$ field to the 
intermediate scale and is therefore undesirable. However, we also expect to
have a renormalizable coupling in the superpotential of the form 
${\bf 16}_H{\bf 10}_H{\bf 16}_H$, which will lead to a term of the form
$f'v_{BL} \chi_d H_u$ below the scale $v_{BL}$ where $\chi_d$ is the
$SU(2)_L$ doublet in the ${\bf 16}_H$. One can now make a change of 
basis and conclude that the linear combination 
$\Sigma_m f_m L_m+f'\chi_d$ becomes super-heavy leaving  the three 
orthogonal combination massless. Those
will be identified with the lepton doublets of the MSSM. This will not induce
any R-parity violating terms in the effective low energy theory. 

Let us however point out that when there are two or more {\bf 10}-dimensional 
multiplets in the theory as is often phenomenologically required, there  
can be mixing between $H_u$ and $H_d$ type doublets. The low energy 
lepton doublets will then acquire components of $H_d$ doublet of MSSM.
This will in turn lead to R-parity violating $QLe^c$ and $LLe^c$ type
terms of strength $m_b/m_{Z}$ or less. Such terms are however absent
if we assume in our discussion that the doublet triplet splitting is 
implemented
by using the Dimopoulos-Wilczek mechanism augmented 
as in Ref.\cite{bm}. In this case,
at the intermediate scale, there are no terms mixing $H_u$ and $H_d$
and hence there no lepton number violating R-parity breaking terms
induced by the renormalizable terms of the superpotential.

Let us now turn to the $\Delta B \neq 0$ terms induced at the high scale.
For this purpose, we have to consider nonrenormalizable
terms in the $SO(10)$ model. They are of the form ${\bf 16}_H {\bf 16}_m 
{\bf 16}_m {\bf 16}_m/ M_{Pl}$. When ${\nu_H^c}$ vev is turned on,
these type of terms lead to terms of type $QLD^c$, $LLE^c$ as well as
$U^cD^cD^c$. Their strength will be given by 
\begin{equation}
\lambda \sim \langle {\nu_H^c}\rangle/
M_{Pl},
\end{equation}
and will therefore depend on the scale of $B-L$ breaking, which
in turn is tied with the gauge coupling unification. Clearly, in single
scale SISUSY GUT models, $\langle{\nu_H}^c\rangle \simeq 2\times 10^{16}$
GeV so that the strength of $\Delta B \neq 0$ terms have strengths of order
$10^{-2}$ to $10^{-3}$, which is too large. On the other hand if there
is an intermediate scale consistent with gauge coupling unification as we
will show below, these couplings get further suppressed by a factor 
of $M_{B-L}/M_{GUT}$
and may be more tolerable if $M_{B-L}$ is small enough. However, in SO(10) GUT 
the quarks and the leptons being in the same GUT multiplet imply that 
L violation occurs whenever there is a B-violation. R-parity violating 
proton decay thus becomes inevitable. To satisfy the proton decay 
constraints \cite{hinchliffe} ($\lambda^\prime \lambda^{\prime \prime} 
\simeq 10^{-24}$) we will require
\begin{equation}
M_{B-L}/M_{GUT} < 10^{-12}-10^{-11}~~~[SO(10) ~case] \label{so10}
\end{equation}  
This will be satisfied if $M_{B-L}\leq 10^{7}$ GeV.
To summarize, we re-examined whether the SO(10) gauge symmetry
alone could suppress the R-parity violating couplings. It turns out 
that if matter parity\footnote{ Under matter parity 
matter fields change sign whereas the Higgs fields do not.} 
is not respected by the 
SO(10) invariant superpotential, the higher dimensional operators 
suppressed by a single power of Planck scale could induce large enough 
R-parity violation even if ${\bf 16_m} \times {\bf 16_m} \times {\bf 16_m}$ 
does not possess an SO(10) singlet. Such induced couplings may lead to 
catastrophic proton decay unless the condition in Eqn. \ref{so10} is 
satisfied.

This leaves us with two possibilities: either we look for theories
where $M_{B-L}\leq 10^7$ GeV or so\footnote{An early example of such
a model was given in Ref.\cite{desh}} 
or alternative groups where proton
decay by itself can be suppressed without eliminating all R-parity
violating couplings. This motivates us to study two 
cases: (i) $G_{224}$ case and (ii) the $[SU(3)]^3$ case\footnote{Similar
arguments also apply to the flipped SU(5) model\cite{lopez}; 
we do not consider this further}.


\noindent{\it $G_{224}$ case:}


In this case, we denote $Q\equiv(2,1,4)$ and $Q^c\equiv (1,2,\bar{4})$
as the matter representations and corresponding Higgs representations
written as $Q_H$, $Q^c_H$ and $\bar{Q}_H$ and $\bar{Q}^c_H$ along with Higgs
fields in the $(1,1,15)$ and $(2,2,0)$ representations. 
The splitting of the GUT multiplets enables one to impose weaker discrete
symmetries so as to have remnant (but not negligible) R-violating interactions
at low energies. As an example, the non-renormalizable terms
that are relevant for our discussion are of the form $Q^cQ^cQ^cQ^c_H$,
$QQQQ$ and $QQQ^cQ^c_H$. If we impose a $Z_4$ symmetry on the theory so
that the last term is forbidden but the the first term is allowed, then
there is no constraint from proton decay and the strengths 
of R-violating $u^cd^cd^c$ terms obtained from the intermediate scale
scenarios with $M_{B-L}\simeq 10^{12}$ or so is in the observationally 
interesting range. They can lead to observable $N-\overline{N}$ 
oscillations if
\begin{equation}
M_{B-L}/M_{GUT} < 10^{-5}
\end{equation}  
A similar situation occurs in the $[SU(3)]^3$ model which 
we shall discuss now.


\noindent{\it  $[SU(3)]^3$ case:}


We will assume the following multiplet structure. The Higgs and matter 
multiplets in this case belong to representations $({\bf 3, 1, 3})$, 
$({\bf 1,\overline{3}, \overline{3}})$ and $({\bf \overline{3}, 3, 1})$ 
representations. The particle content of the bi-triplet
representations can be given by:
\begin{eqnarray}
\pmatrix{ u \cr d \cr g }
\pmatrix{ u^c \cr d^c \cr g^c }
\pmatrix{
H^0_u & H^+_d & e^+ \cr
H^-_u & H^0_d & \nu^c \cr
e^-   & \nu   & n^0 
} \label{fields}
\end{eqnarray}

In order to investigate the nature of R-parity breaking in this model, let
us use the notation where $\psi$,  $\psi^c$ and $\lambda$ denote the
above three 
representations respectively. There can be Higgs superfields of type 
$\lambda$ and $\bar{\lambda}$; we will denote them with a subscript $H$.
An important point to remember is that there are no matter multiplets 
of $\overline{\lambda}$ type. The matter fields will have either no 
subscript or the generation index where needed. 
There are several types of gauge invariant renormalizable  couplings - 
they are: $\psi \psi^c \lambda$, $\psi^3$, ${\psi^c}^3$, $\lambda^3$, 
$\bar{\lambda}^3$, $\lambda \bar{\lambda}_H$ etc. There are also higher 
dimensional string or Planck scale induced terms. A realistic 
$[SU(3)]^3$ model with doublet-triplet splitting requires at least 
two pair of $\lambda_H$+$\bar{\lambda}_H$\cite{shafi}. We will work 
within the framework of such a model when attempting to make contact 
with low energy physics.

Let us first note that in the symmetry limit, the models conserve R-parity
due to the presence of $B-L$ as part of the gauge symmetry. However, the 
R-parity violating terms arise once the $\nu_H^c$ vev is inserted in the 
above operators. There can be various sources for the R-parity breaking 
terms. It then follows that any R-parity violating interaction arising 
from induced non-renormalizable term will have strength of order $M_{B-L}/M_{GUT}$ 
or so. For the sake of illustration, we will focus on the induced
$\Delta B\neq 0$ term below. 
 
Using the notations of Eq. \ref{fields}, we write down the R-parity 
violating  (B-violating) couplings \cite{dis} which always involve 
at-least one exotic particle.
\begin{equation}
{\cal W}_B= \lambda_1~u^c ~d^c ~g^c + \lambda_2 ~ u ~d ~g
\end{equation}
The vev of $n_H^0$ gives the direct mass term for $g$ and $g^c$ whereas 
that of $\nu_H^c$ mixes the fields $g^c$ and $d^c$ by the term
\begin{equation}
{\cal W}_{mass}= \lambda_3~g~g^c~n_H^0~~;~~{\cal W}_{mix}=\lambda_4
~d^c~g~\nu_H^c
\end{equation}
We are lead to the mass matrix of the d-type quarks of the form,
\[
\begin{array}{c c}
& 
\begin{array}{c c}
d^c~~~&~~~~~g^c
\end{array} \\
\begin{array}{c}
d \\ g
\end{array} &
\left( \begin{array}{c c}
m_d & 0 \\
\lambda_4~\langle \nu_H^c \rangle & \lambda_3~\langle n_H^0 \rangle
\end{array} \right)
\end{array}
\]

\begin{figure}[h]
\begin{center}
\epsfxsize=8.5cm
\epsfysize=8.5cm
\mbox{\hskip -1.0in}\epsfbox{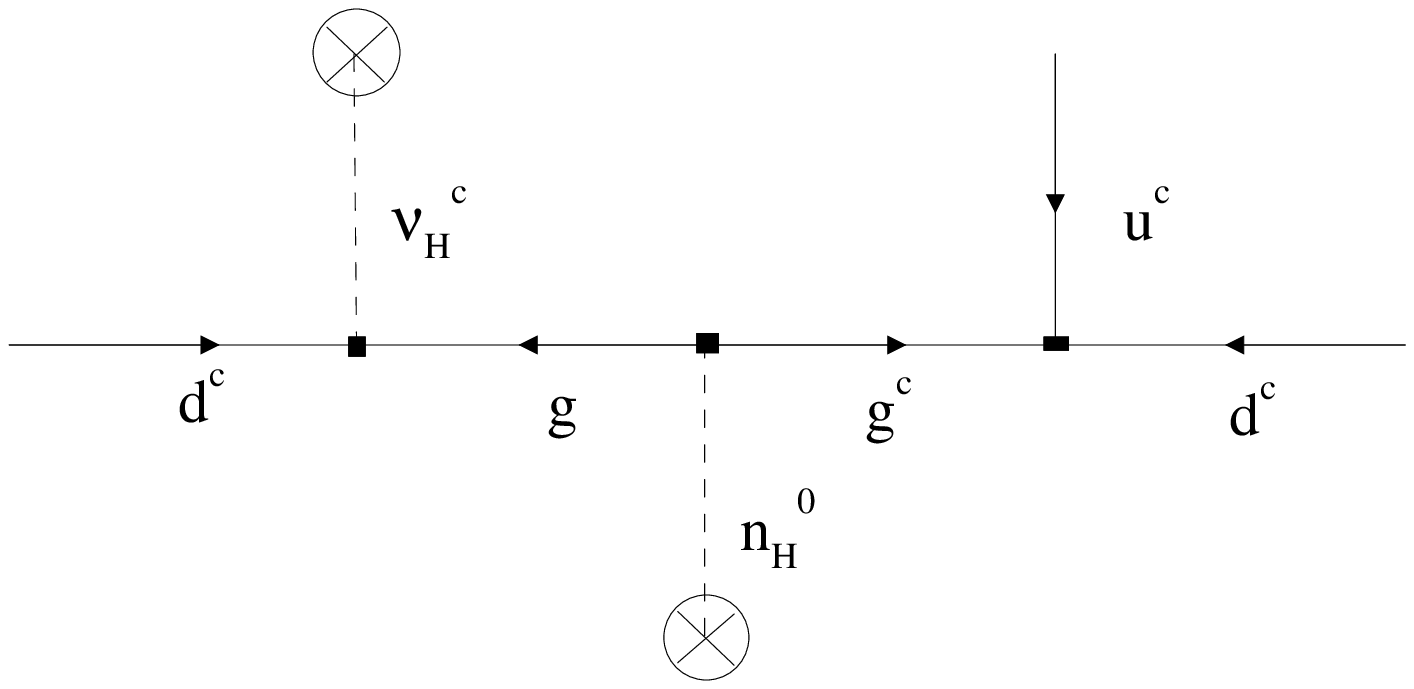}
\caption{ The graph generating the $u^c d^c d^c$ effective vertex.
\label{fig1}
}
\end{center}
\end{figure}

The B-violating couplings are generated by the diagram given in Fig. 
\ref{fig1} and their strengths can be estimated to be, 
\begin{equation}
\lambda^{\prime \prime}={\lambda_1 \lambda_4 \over \lambda_3}
{\langle \nu_H^c\rangle \over {\langle n_H^0\rangle}}.
\end{equation} 
The other coupling involving $\lambda_4$ is still much smaller \cite{dis}. 
Thus we see that the strength of $\Delta B\neq 0$
R-parity violating terms are dictated by gauge coupling unification
and their suppression depends crucially on the $B-L$ breaking scale
being an intermediate scale. We should also note that in the absence of the
discrete symmetry  $L \rightarrow -L$ the L violating couplings can also be
generated by an identical mechanism. By the arguments given in the
$SO(10)$ case above, the presence of the L-violating terms would to
rapid proton decay unless the corresponding couplings are suppressed to
the level of $10^{-12}$ which is possible only in the case where the
$M_{B-L}\simeq 10^{6}$ leaving no testable 
prediction in the B- and L-violating 
sector. We will therefore impose the above $ L \rightarrow -L$ symmetry. 
In this case (as in the $G_{224}$ case above), a higher intermediate
scale could have interesting experimental implications via enhanced
R-parity violating couplings.

 Let us therefore proceed to discuss 
gauge coupling unification constraints on the scales of $B-L$ breaking 
and $M_{GUT}$ in both the $[SU(3)]^3$ and SO(10) models.

\begin{table}[t]
\begin{center}
\[
\begin{array}{|c||c|c|c|c|c|}
\hline
Model & n_L & n_R & n_H & n_C &n_d  \\
\hline
I & 0& 1& 2& 1&1 \\
II& 1& 1& 1& 1 &1\\ 
III & 1 & 3 & 2 & 1 & 2 \\
\hline
\end{array}
\]
\end{center}
\caption{Various models with particle contents above the intermediate scale.
The symbols $n_L,~ n_R,~n_H,~n_c$ denote the number of 
fields above the intermediate scale with quantum numbers under the
gauge group $SU(3)_c\times SU(2)_L\times SU(2)_R\times U(1)_{B-L}$ as follows: 
$n_L\equiv (1,2,1,1)+(1,2,1,-1)$, $n_R\equiv (1,1,2,1)+(1,1,2,-1)$,
$n_H\equiv (1,2,2,0)$, $n_c\equiv (8,1,1,0)$. The symbol $n_d$ denotes
the number of pairs of MSSM type Higgs doublets below $M_I$.}
\end{table}

\section{Gauge coupling unification and scale of $B-L$ breaking}

\begin{figure}[htb]
\begin{center}
\epsfxsize=8.5cm
\epsfysize=6.5cm
\mbox{\hskip -1.0in}\epsfbox{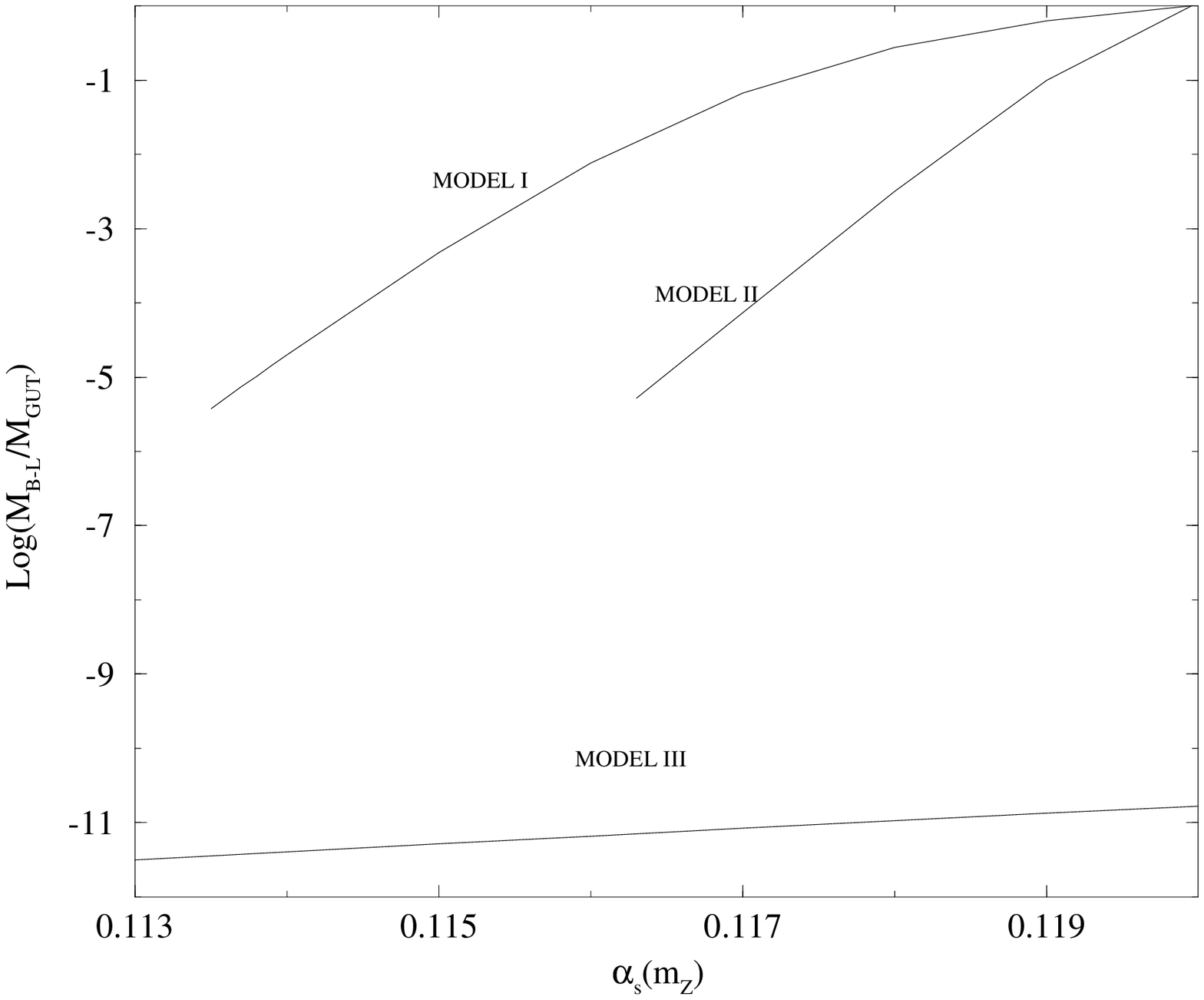}
\caption{ The two-loop predictions of $M_{B-L} \over M_{GUT}$.
\label{fig2}
}
\end{center}
\end{figure}

In order to discuss the gauge coupling unification
and the constraints on the $B-L$ breaking scale in these models, we 
assume that the symmetry 
in the intermediate scale corresponds to $SU(3)_c\times SU(2)_L
\times SU(2)_R\times U(1)_{B-L}$. For the $[SU(3)]^3$ model,
this means that the GUT symmetry is broken by the vev of the $n^0$
field. The spectrum of particles below $M_{GUT}$ can be assumed to be same as that
for the $SO(10)$ case. Therefore, only one set of discussions given below applies 
to both the cases. The two loop evolution of gauge couplings in the intermediate scale 
unified models \cite{brahm1} has been performed. The Higgs 
contents of the models are given in Table 1. The models are chosen in 
such a way that the splitting between the 
unification scale and the intermediate
scale is either of the order of $10^5$ GeV or of the order of $10^{12}$ GeV 
and also the unification scale is around the
string scale where the GUT symmetry breaks to the 
left-right symmetry\footnote{For recent models with $M_{GUT}\simeq M_{string}$
and $M_I\simeq 10^{12}~GeV$, see\cite{brahm1,dutta}}.
The results are summarized in Table 2. 
A plot of the variation of the ratio $M_{B-L} 
\over M_{GUT}$ with respect to $\alpha_s(m_Z)$ is given in Fig. \ref{fig2}.

\begin{table}[t]
\begin{center}
\[
\begin{array}{|c|c|c|c|c|}
\hline
Model& \alpha_s& M_{B-L}& M_{GUT}& \lambda \\
\hline
I   & 0.1135 & 10^{13.6} & 10^{19}  & 10^{-5.4}\\
II  & 0.1163 & 10^{13.7} & 10^{19}  & 10^{-5.3}\\
III & 0.1130 & 10^{6.9}  & 10^{18.5}& 10^{-11.5}\\
\hline
\end{array}
\]
\end{center}
\caption{The mass scales and the predicted values of the R-parity
violating couplings.}
\end{table}

Model III has a splitting of order $10^{12}$ between the intermediate
scale and the GUT scale with a very mild variation with respect to 
$\alpha_s(m_Z)$. Consequently the R-parity violating couplings
will be of order $10^{-12}$. In the case of SO(10), where it is not possible 
to distinguish the leptons and the baryons by a discrete symmetry of the type
(L$\rightarrow$-L), the baryon and lepton number violation must coexist and
in such a case R-parity violating couplings of order $10^{-12}$ are required
to suppress the R-parity violating proton decay. The models I and II are
good in the case when the GUT symmetry is $G_{224}$ or $[SU(3)]^3$ and a 
GUT level leptonic parity can be imposed. Note that the price we pay for 
case III is that the model below $M_{B-L}$ is not MSSM but MSSM with an extra 
pair of Higgs doublets. When we insist on recovering MSSM below $M_{B-L}$ we 
do not get the interesting case of $M_{GUT}=M_{string}$. 
We note that to increase the unification scale up to the string  scale we had 
to introduce a color octet Higgs field at the intermediate scale. To keep 
the octet at the intermediate scale we need to do some fine tuning of the
Higgs potential at the GUT scale. 

\section{Observable neutron-anti-neutron oscillation}

 We see from the discussion in the above sections that operators 
of type $u^c d^c d^c$ are induced with strength of order $\lambda f$
where $\lambda \simeq 10^{-4}$ as determined by the unification analysis
and $f={\lambda_1 \lambda_4 \over \lambda_3}$ is an unknown parameter (which could be 
assumed to be of order $10^{-1}$). It can lead to neutron-anti-neutron 
oscillation  since it is a six-quark operator $u^cd^cd^cu^cd^cd^c$ in terms of 
superfields. The strength of this operator naively is $\lambda^2f^2$.
As is well-known, the above six-quark operator in the supersymmetric case
is non-zero only when it connects two different generations. 
Therefore
there is an extra suppression arising from quark mixing. It has been
shown by Goity and Sher\cite{goity} that the lowest order in which this operator
contributes to $N-\overline{N}$ oscillation involves a box diagram
involving the wino and leads to the strength for the six fermion $N-
\overline{N}$ operator:
\begin{eqnarray}
\lambda_{N-\overline{N}}\simeq {{3g^4 \lambda^2f^2 A m_b m_{
\tilde{w}}}V^2_{ub}
\over{8\pi^2 M^4_{\tilde{b}_L} M^4_{\tilde{b}_R}}} 
\end{eqnarray}
In order to estimate the transition time for neutron-anti-neutron oscillation,
we have to multiply by the wave function effect i.e. $|\psi(0)|^2$:
\begin{eqnarray}
\tau^{-1}_{N-\overline{N}}=\lambda_{N-{\overline{N}}} |\psi(0)|^2
\end{eqnarray}
Using the value for $|\psi(0)|^2\simeq 3\times 10^{-4}$ from Ref.\cite{pasu},
we get

\begin{eqnarray}
\tau^{-1}_{N-\overline{N}}\simeq 5\times 10^{-25} \lambda^2f^2 \left({{300 GeV}
\over{M_{sq}}}\right)^6
\end{eqnarray}

The unification analysis of the previous section implies that $\lambda\simeq
10^{-4}$ implying an $N-\overline{N}$ oscillation time of $10^8f^{-2}$ sec.,
whereas the present experimental lower limit on this process is 
$10^8$ sec.\cite{baldo}. Thus we see that such superstring inspired
supersymmetric models can be tested by the neutron-anti-neutron oscillation
or by $\Delta B=2$ proton decay models such as $N+P\rightarrow n \pi$. We
hasten to note that due to the unknown coupling $f$ in the six-quark superfield
operator, we cannot make an exact prediction; but we expect the prediction 
for the neutron-anti-neutron oscillation time to
be somewhere between $10^8$ to $10^{10}$ sec. There is a recent proposal
by a group at Oak Ridge National laboratory to search for neutron-anti-neutron
oscillation up to a sensitivity of $10^{10}$ to $10^{11}$ sec.\cite{yuri} 
which should therefore throw light on the nature of this class of grand 
unified theories.

 
\section{Conclusion}


In conclusion, we have shown that  superstring inspired SUSY GUT
models, based on the gauge groups $SO(10)$ and $[SU(3)]^3$, 
can lead to large R-parity breaking interactions of $\Delta L
\neq 0$ and $\Delta B\neq 0$ type after symmetry breaking unless the
$B-L$ gauge symmetry breaks at an intermediate scale. 
We further show that for the $SO(10)$ model $M_{B-L}$ is not 
low enough, both L and 
B-violating couplings of comparable magnitude are induced by higher 
dimensional operators. Thus, SO(10) gauge symmetry is not enough to 
suppress the R-parity violating proton decay. The SO(10) scenarios where 
intermediate scale is not small enough to have acceptable R-violation are 
forced to be completely R-conserving by some additional symmetry as 
matter parity. 
For the $G_{224}$ and $[SU(3)]^3$ case however, it is 
possible to impose additional symmetries that are different from R-parity 
and obtain MSSM at low energies and yet have acceptable R-violating terms.
Explicit realization of such intermediate scale scenarios are given.
We further show that for the $[SU(3)]^3$ and $G_{224}$ case,
one can have observable neutron-anti-neutron oscillation
with a transition time in the range $10^{10}$ sec. or so.

\section{Acknowledgments} B. B. acknowledges discussions 
with A. Yu. Smirnov and F. Vissani.


\begin{thebibliography}{99}
\bibitem{dis} $[SU(3)]^3$ can be decomposed to the basis
$SU(3)_c \times SU(2)_L \times U(1) \times U(1) \times U(1)$. It is a
convenient basis to check that these are the only two baryon-number
violating terms. The mixing between the d and g is of the order
$(\langle \nu_H^c \rangle / \langle n_H^0 \rangle) (m_d /\langle 
n_H^0 \rangle)$.
\end{thebibliography}
\end{document}